\documentclass{emulateapj} \usepackage{apjfonts}

\newcommand{\magiicat}{\hbox{{\rm MAG}{\sc ii}CAT}}

\newcommand{\etal}{et~al.}
\newcommand{\PVdblt}{{\rm P}\kern 0.1em{\sc v}~$\lambda\lambda 1117, 1128$}
\newcommand{\CaIIdblt}{{\rm Ca}\kern 0.1em{\sc ii}~$\lambda\lambda 3934, 3969$}
\newcommand{\AlIIIdblt}{{\rm Al}\kern 0.1em{\sc iv}~$\lambda\lambda 1855, 1863$}
\newcommand{\CIVdblt}{{\rm C}\kern 0.1em{\sc iv}~$\lambda\lambda 1548, 1550$}
\newcommand{\MgIIdblt}{{\rm Mg}\kern 0.1em{\sc ii}~$\lambda\lambda 2796, 2803$}
\newcommand{\NVdblt}{{\rm N}\kern 0.1em{\sc v}~$\lambda\lambda 1238, 1242$}  
\newcommand{\SVIdblt}{{\rm S}\kern 0.1em{\sc vi}~$\lambda\lambda 933, 944$} 
\newcommand{\OVIdblt}{{\rm O}\kern 0.1em{\sc vi}~$\lambda\lambda 1031, 1037$} 
\newcommand{\SiIIdblt}{{\rm Si}\kern 0.1em{\sc ii}~$\lambda\lambda 1190, 1193$} 
\newcommand{\SiIVdblt}{{\rm Si}\kern 0.1em{\sc iv}~$\lambda\lambda 1393, 1402$} 
\newcommand{\PV}{\hbox{{\rm P}\kern 0.1em{\sc v}}}
\newcommand{\AlI}{\hbox{{\rm Al}\kern 0.1em{\sc i}}}
\newcommand{\AlII}{\hbox{{\rm Al}\kern 0.1em{\sc ii}}}
\newcommand{\AlIII}{{\hbox{\rm Al}\kern 0.1em{\sc iii}}}
\newcommand{\CaII}{\hbox{{\rm Ca}\kern 0.1em{\sc ii}}}
\newcommand{\CII}{\hbox{{\rm C}\kern 0.1em{\sc ii}}}
\newcommand{\CIIe}{\hbox{{\rm C$^{\ast}$}\kern 0.1em{\sc ii}}}
\newcommand{\CIII}{\hbox{{\rm C}\kern 0.1em{\sc iii}}}
\newcommand{\CIV}{\hbox{{\rm C}\kern 0.1em{\sc iv}}}
\newcommand{\CV}{\hbox{{\rm C}\kern 0.1em{\sc v}}}
\newcommand{\HI}{\hbox{{\rm H}\kern 0.1em{\sc i}}}
\newcommand{\HII}{\hbox{{\rm H}\kern 0.1em{\sc ii}}}
\newcommand{\Lya}{\hbox{{\rm Ly}\kern 0.1em$\alpha$}}
\newcommand{\Lyb}{\hbox{{\rm Ly}\kern 0.1em$\beta$}}
\newcommand{\Lyg}{\hbox{{\rm Ly}\kern 0.1em$\gamma$}}
\newcommand{\Lyd}{\hbox{{\rm Ly}\kern 0.1em$\delta$}}
\newcommand{\Lye}{\hbox{{\rm Ly}\kern 0.1em$\epsilon$}}
\newcommand{\Lyphi}{\hbox{{\rm Ly}\kern 0.1em$\phi$}}
\newcommand{\Lyfive}{\hbox{{\rm Ly}\kern 0.1em$5$}}
\newcommand{\Lysix}{\hbox{{\rm Ly}\kern 0.1em$6$}}
\newcommand{\Lyseven}{\hbox{{\rm Ly}\kern 0.1em$7$}}
\newcommand{\Lyeight}{\hbox{{\rm Ly}\kern 0.1em$8$}}
\newcommand{\Lynine}{\hbox{{\rm Ly}\kern 0.1em$9$}}
\newcommand{\Lyten}{\hbox{{\rm Ly}\kern 0.1em$10$}}
\newcommand{\Lyeleven}{\hbox{{\rm Ly}\kern 0.1em$11$}}
\newcommand{\HeI}{\hbox{{\rm He}\kern 0.1em{\sc i}}}
\newcommand{\HeII}{\hbox{{\rm He}\kern 0.1em{\sc ii}}}
\newcommand{\FeI}{\hbox{{\rm Fe}\kern 0.1em{\sc i}}}
\newcommand{\FeII}{\hbox{{\rm Fe}\kern 0.1em{\sc ii}}}
\newcommand{\FeIII}{\hbox{{\rm Fe}\kern 0.1em{\sc iii}}}
\newcommand{\MnII}{\hbox{{\rm Mn}\kern 0.1em{\sc ii}}}
\newcommand{\MgI}{\hbox{{\rm Mg}\kern 0.1em{\sc i}}}
\newcommand{\MgIb}{\hbox{{\rm Mg}\kern 0.1em{\sc i}}\kern 0.05em{\rm b}}
\newcommand{\MgII}{\hbox{{\rm Mg}\kern 0.1em{\sc ii}}}
\newcommand{\MgIII}{\hbox{{\rm Mg}\kern 0.1em{\sc iii}}}
\newcommand{\NI}{\hbox{{\rm N}\kern 0.1em{\sc i}}}
\newcommand{\NII}{\hbox{{\rm N}\kern 0.1em{\sc ii}}}
\newcommand{\NIII}{\hbox{{\rm N}\kern 0.1em{\sc iii}}}
\newcommand{\NV}{\hbox{{\rm N}\kern 0.1em{\sc v}}}
\newcommand{\OVI}{\hbox{{\rm O}\kern 0.1em{\sc vi}}}
\newcommand{\OI}{\hbox{{\rm O}\kern 0.1em{\sc i}}}
\newcommand{\OII}{\hbox{[{\rm O}\kern 0.1em{\sc ii}]}}
\newcommand{\OIII}{\hbox{[{\rm O}\kern 0.1em{\sc iii}]}}
\newcommand{\OIV}{\hbox{{\rm O}\kern 0.1em{\sc iv}]}}
\newcommand{\SI}{{\rm S}\kern 0.1em{\sc i}}
\newcommand{\SIV}{{\rm S}\kern 0.1em{\sc iv}}
\newcommand{\SVI}{{\rm S}\kern 0.1em{\sc vi}}
\newcommand{\SiI}{\hbox{{\rm Si}\kern 0.1em{\sc i}}}
\newcommand{\SiII}{\hbox{{\rm Si}\kern 0.1em{\sc ii}}}
\newcommand{\SiIII}{\hbox{{\rm Si}\kern 0.1em{\sc iii}}}
\newcommand{\SiIV}{\hbox{{\rm Si}\kern 0.1em{\sc iv}}}
\newcommand{\SII}{\hbox{{\rm S}\kern 0.1em{\sc ii}}}
\newcommand{\SIII}{\hbox{{\rm S}\kern 0.1em{\sc iii}}}
\newcommand{\NaI}{\hbox{{\rm Na}\kern 0.1em{\sc i}}}
\newcommand{\NaID}{\hbox{{\rm Na}\kern 0.1em{\sc i}}\kern 0.05em{\rm D}}
\newcommand{\TiII}{\hbox{{\rm Ti}\kern 0.1em{\sc ii}}}
\newcommand{\kms}{\hbox{km~s$^{-1}$}}
\newcommand{\cmsq}{\hbox{cm$^{-2}$}}

\slugcomment{ApJL: Accepted Sept. 26, 2013} 
\shorttitle{\sc $D<6$~kpc {\MgII} Absorbers}
\shortauthors{\sc Kacprzak et~al.}

\begin{document}


\title{The Smooth {\MgII} gas distribution through the
  interstellar/extra-planar/halo interface}


\author{\sc
Glenn G. Kacprzak\altaffilmark{1,2},
Jeff Cooke\altaffilmark{1}, 
Christopher W. Churchill\altaffilmark{3},
Emma V. Ryan-Weber\altaffilmark{1},
and
Nikole M. Nielsen\altaffilmark{3}
}
                                                                                
\altaffiltext{1}{Swinburne University of Technology, Victoria 3122,
Australia {\tt gkacprzak@astro.swin.edu.au}}
\altaffiltext{2}{Australian Research Council Super Science Fellow}
\altaffiltext{3}{New Mexico State University, Las Cruces, NM 88003}

\begin{abstract}

We report the first measurements of {\MgII} absorption systems
associated with spectroscopically confirmed $z\sim0.1$ star-forming
galaxies at projected distances of $D<6$~kpc.  We demonstrate the data
are consistent with the well known anti-correlation between rest-frame
{\MgII} equivalent width, $W_r(2796)$, and impact parameter, $D$,
represented by a single log-linear relation derived by Nielsen et
al. (\magiicat) that converges to $\sim2$~{\AA} at $D=0$~kpc.
Incorporating \magiicat, we find that the halo gas covering fraction
is unity below $D\sim25$~kpc. We also report that our $D<6$~kpc
absorbers are consistent with the $W_r(2796)$ distributions of the
Milky Way interstellar medium (ISM) and ISM+halo.  In addition, quasar
sight-lines of intermediate redshift galaxies with $6<D<25$~kpc have
an equivalent width distribution similar to that of the Milky Way
halo, implying that beyond $\sim6$~kpc, quasar sight-lines are likely
probing halo gas and not the ISM.  As inferred by the Milky Way and
our new data, the gas profiles of galaxies can be fit by a single
log-linear $W_r(2796)-D$ relation out to large scales across a variety
of gas-phase conditions and is maintained through the
halo/extra-planar/ISM interfaces, which is remarkable considering
their kinematic complexity.  These low redshift, small impact
parameter absorption systems are the first steps to bridge the gap
between quasar absorption-line studies and {\HI} observations of the
CGM.

\end{abstract}




\section{Introduction}
\label{sec:intro}

\begin{center}
\begin{deluxetable*}{llcrrcrrllc}
\tabletypesize{\scriptsize} \tablecaption{Keck--I/LRIS Quasar/Galaxy
Observations\label{tab:LRIS}} \tablecolumns{10} \tablewidth{0pt}

\tablehead{
\colhead{ID}&
\colhead{SDSS}&
\colhead{$z_{{\rm qso}}$}&
\colhead{RA} &
\colhead{DEC} &
\colhead{$\theta$ } &
\colhead{$D$ } &
\colhead{$z_{{\rm gal}}$} &
\colhead{$z_{{\rm abs}}$}&
\colhead{$W_r(2796)$} &
\colhead{$W_r(2803)$} \\
\colhead{ } &
\colhead{Quasar Name}&
\colhead{ } &
\colhead{(J2000)} &
\colhead{(J2000)} &
\colhead{($''$)} &
\colhead{(kpc)} &
\colhead{ } &
\colhead{ } &
\colhead{(\AA)} &
\colhead{(\AA)} 
}
\startdata
12  & J100514.20$+$530240.0& 0.561& 10:05:14.19 & $+$53:02:40.39 & $0.98\pm0.04$ & $2.36\pm0.10$  &0.1358  &0.135215& 2.46$\pm$0.16  &  2.24$\pm$0.14   \\
16  & J110735.68$+$060758.6& 0.380& 11:07:35.68 & $+$06:07:58.62 & $1.52\pm0.01$ & $4.06\pm0.03$  &0.1545  &0.154217& 3.11$\pm$0.32  &  2.61$\pm$0.31   \\
19  & J123844.79$+$105622.2& 1.304& 12:38:44.79 & $+$10:56:22.23 & $0.55\pm0.02$ & $1.18\pm0.04$  &0.1185  &0.116537& 1.75$\pm$0.53  &  2.51$\pm$0.40   \\
17  & J143458.04$+$504118.6& 1.485& 14:34:58.06 & $+$50:41:18.93 & $1.61\pm0.09$ & $5.30\pm0.30$  &0.1992  &0.198697& 2.94$\pm$0.20  &  2.91$\pm$0.19   \\
14  & J145240.54$+$544345.3& 1.519& 14:52:40.54 & $+$54:43:45.34 & $0.62\pm0.10$ & $1.17\pm0.19$  &0.1020  &0.101963& 2.13$\pm$0.49  &  2.33$\pm$0.41   \\
21  & J145938.49$+$371314.7& 1.219& 14:59:38.49 & $+$37:13:14.69 & $1.31\pm0.09$ & $3.40\pm0.23$  &0.1489  &0.148484& 2.16$\pm$0.24  &  1.93$\pm$0.20   \\
7   & J160521.27$+$510740.8& 1.229& 16:05:21.27 & $+$51:07:41.29 & $2.12\pm0.08$ & $3.90\pm0.15$  &0.0994  &0.098006& 2.35$\pm$0.38  &  2.68$\pm$0.22   
\enddata
\end{deluxetable*}  
\end{center}

Metal enrichment and the interplay between galaxies and their
extra-planar/halo gas is strongly influenced by mergers, galactic
winds, high velocity clouds (HVCs), and filamentary infall \citep[see
  reviews by][]{putman12,sancisi08,veilleux05}.  However, we lack a
thorough understanding of how these processes affect galaxies, their
extra-planar gas, and their surrounding circumgalactic medium (CGM).
Observation of the quantity, distribution and properties of gas within
galaxies and their halos can provide vital insight into the physical
processes that govern the dynamics and chemical enrichment of galaxies
and the CGM.

The {\MgIIdblt} absorption doublet, detected in background quasar
spectra and known to arise from the gaseous halos of galaxies, is an
ideal tracer of low-ionization gas with $10^{16}\leq N(\HI)\leq
10^{22}$~{\cmsq} \citep{archiveI,weakII,cwc-china}. The combination of
the quasar absorption-line technique and a sensitive CGM tracer
provides a unique means of directly observing the interplay and
mechanisms by which galaxies acquire, expel, chemically enrich, and
recycle their gaseous component. Furthermore, a significant quantity
of {\HI} is probed by {\MgII} absorption; roughly equivalent to 5\% of
the total hydrogen in stars \citep{kacprzak11c,menard12}.


Although there is significant evidence suggesting that {\MgII}
absorption traces outflows from star-forming galaxies
\citep{bouche06,tremonti07,zibetti07,noterdaeme10,
  rubin10,bordoloi11,menard12,martin12,rubin13}, accretion onto
galaxies \citep{steidel02,chen10a,kacprzak10a,kacprzak11b,
  ribaudo11,churchill12,kacprzak12,martin12,churchill13a,rubin13} and
high velocity clouds \citep{richter12}, with all indicating
inclination and azimuthal angle dependencies
\citep{kacprzak11b,kacprzak12,bordoloi11,bordoloi12,bouche12}, the
      {\MgII} rest-frame equivalent width is strongly anti-correlated
      with impact parameter \citep[e.g.,][]{steidel95,
        bouche06,kacprzak08,chen10a,churchill13a,nielsen12,nielsen13}.

With 182 galaxies having impact parameters of $D\gtrsim10~$kpc,
\citet{nielsen12,nielsen13} showed a 7.9~$\sigma$ anti-correlation
between $W_r(2796)$ and $D$. However, it is unclear how this
relationship behaves below $D\sim10$~kpc given that within this regime
resides the boundaries between the extended halo gas, the extra-planar
gas, HVCs, and the interstellar medium of the host galaxy. Although
the $W_r(2796)-D$ relation is smooth, there exists considerable
scatter in the data that could be related to the host galaxy
luminosity, mass, star formation, orientation, or to a patchy CGM
distribution \citep[see][and references
  therein]{nielsen12,churchill13a,churchill13b}.  At low impact
parameters, the gas covering fraction is expected to be unity
\citep{nielsen13} since local {\HI} observations show a unity gas
covering fraction within a few kiloparsecs \citep{zwaan05b}.

In the local Universe, there is a wealth of {\HI} emission-line
studies examining the complex extra-planar gas dynamics within 15~kpc
of galaxies (limited to logN(\HI)$<19$)
\citep{putman12,sancisi08,veilleux05}.  Given that the majority of the
gas physics occurs near/above the plane of the disk, it is critical to
identify absorption-line systems at low $D$ to examine how the
$W_r(2796)-D$ relation behaves around the disk/extra-planar/halo gas
interface. Furthermore, absorption systems at $z\sim0.1$
\citep[e.g.,][]{barton09} will provide overlapping and complementary
data for future {\HI} surveys using The Australian Square Kilometer
Array Pathfinder, such as WALLABY ($600$k galaxies at $z=0-0.25$) and
DINGO ($100$k galaxies at $z=0-0.40$).

Motivated by the lack of overlap between {\HI} emission and
absorption-line system observations and the lack of absorption-line
systems within this low impact parameter range, we identified seven
new {\MgII} absorbers with $D<6$~kpc. These observations provide the
first glimpses into the galaxy/extra-planar/halo gas boundaries
connecting {\HI} and {\MgII} observations.  In \S~\ref{sec:data} we
describe our sample selection and analysis.  In \S~\ref{sec:results},
we present the $D<6$~kpc absorbers and how they fit on the existing
$W_r(2796)-D$ anti-correlation. We further compare our systems to
those in our own Milky Way in order to interpret the plausible origin
of the absorbing gas at very small impact parameters. Our concluding
remarks are in \S~\ref{sec:conclusion}.  We adopt a $h=0.70$,
$\Omega_{\rm M}=0.3$, $\Omega_{\Lambda}=0.7$ cosmology.

\section{The Sample and Analysis}\label{sec:data}

The Sloan Digital Sky Survey (SDSS) spectroscopic fibers have a
diameter of 3$''$ (8~kpc at $z=0.15$), thus, any fiber could contain a
background quasar and a foreground galaxy. Therefore close
star-forming galaxy and quasar pairs can be identified by quasar
spectra having superimposed foreground galaxy emission-line spectra.
\citet{york12} identified 23 H$\alpha$-emitting foreground galaxies
with $z<0.4$ at small impact parameters from quasars using the SDSS
Data Release 5 and \citet{noterdaeme10} identified 46 {\OIII}-emitting
galaxies at $z<0.8$ within the SDSS Data Release 7. Here we have
identified a subset of seven $z\sim0.1$ galaxies that are clearly
identified in ground-based imaging and are at a redshift where it is
possible to detect their {\MgII} absorption using blue sensitive
optical CCDs \citep[see][]{barton09}.

\subsection{Quasar Spectroscopy}

The seven quasar spectra were obtained on 2013 April 10 using the Keck
Low Resolution Imaging Spectrometer (LRIS) \citep{oke95} with the 1200
lines/mm grism blazed at 3400~{\AA}, which covers a wavelength range
of 2910$-$3890\AA. We used a 0.7$''$ slit (1$''$ slit for ID7),
providing a dispersion of 0.17~{\AA} per pixel and a resolution of
$\sim$1.12~{\AA} ($\sim$105~\kms). Integration times of 2200--4400
seconds were used, depending on the magnitude of the quasar and the
foreground galaxy redshift. The spectra were reduced using the
standard IRAF packages and were heliocentric and vacuum corrected.

The quasar spectra were objectively searched for {\MgII} doublet
candidates using a detection significance level of 3~$\sigma$
(2~$\sigma$ for ID19) for the $\lambda$2796 and $\lambda$2803
lines. All seven absorbers are also identifiable by their strong
{\MgI} $\lambda$2853 absorption. Detection and significance levels
follow the formalism of \citet{weakI}.  Analysis of the absorption
profiles was performed using our own graphic-based interactive
software that uses the direct pixel values to measure the equivalent
widths and the redshift of the {\MgII} $\lambda$2796 transition.  The
{\MgII} absorption redshifts were computed from the optical depth
weighted mean of the absorption profiles \citep[see][]{cv01}. The
statistical redshift uncertainties range between 0.00001--0.00009
($\sim3-30$~{\kms} co-moving).

\begin{figure}
\includegraphics[angle=0,scale=0.16]{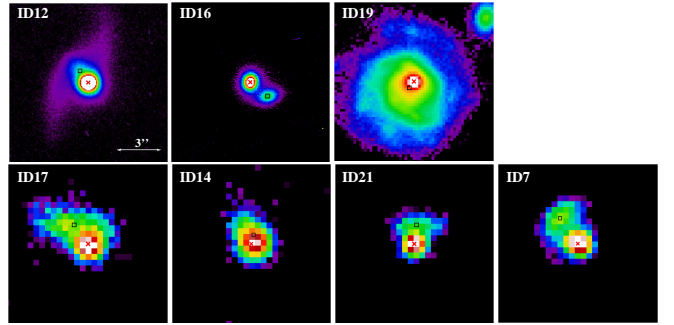}
\caption[angle=0]{15$\times$15$''$ images ($27.7\times 27.7$~kpc at
  $z=0.1$) of the obscured quasar (center indicated by an ``X'') and
  the emission-line foreground galaxies (centers indicated by an
  square) producing the observed {\MgII} absorption.  The close galaxy
  and quasar pairs were identified by the quasar spectra having
  superimposed foreground galaxy emission-line spectra. The top panels
  are Gemini/GMOS-N $i$-band, Gemini/GMOS-S $i$-band and
  CFHT/MegaPrime $g$-band images (left to right). The bottom panels
  are SDSS $r$-band images.  
}
\label{fig:gals} 
\end{figure}

\subsection{Galaxy/Quasar Imaging}

In Figure~\ref{fig:gals}, we present the seven new quasar-galaxy pairs
where the quasar is placed in the center of the image. All galaxies
were imaged with SDSS (0.4$''$/pixel), however additional archival
imaging was obtained for three of the objects from the Canadian
Astronomy Data Centre (CADC). ID12 was imaged on 20 December 2008 for
1100 seconds using Gemini/GMOS-N (0.073$''$/pixel) in the $i$-band
(PID GN-2008B-Q-128), ID16 was imaged on 27 December 2008 for 1100
seconds using Gemini/GMOS-S (0.073$''$/pixel) in the $i$-band (PID
GS-2008B-Q-79) and ID19 was imaged on 25 May 2009 for 3171 seconds
using CFHT/MegaPrime (0.19$''$/pixel) in the $g$-band (PID 09AP03).

Impact parameters were computed between the photometric centroids of
the background quasar and galaxy using Source Extractor
\citep{bertin96}. In cases where the quasar and galaxy are blended
(e.g., ID14), we performed a PSF subtraction of the quasar light. The
PSFs were created by modeling selected stars within the same field.

\begin{figure*}
\begin{center}
\includegraphics[angle=0,scale=0.77]{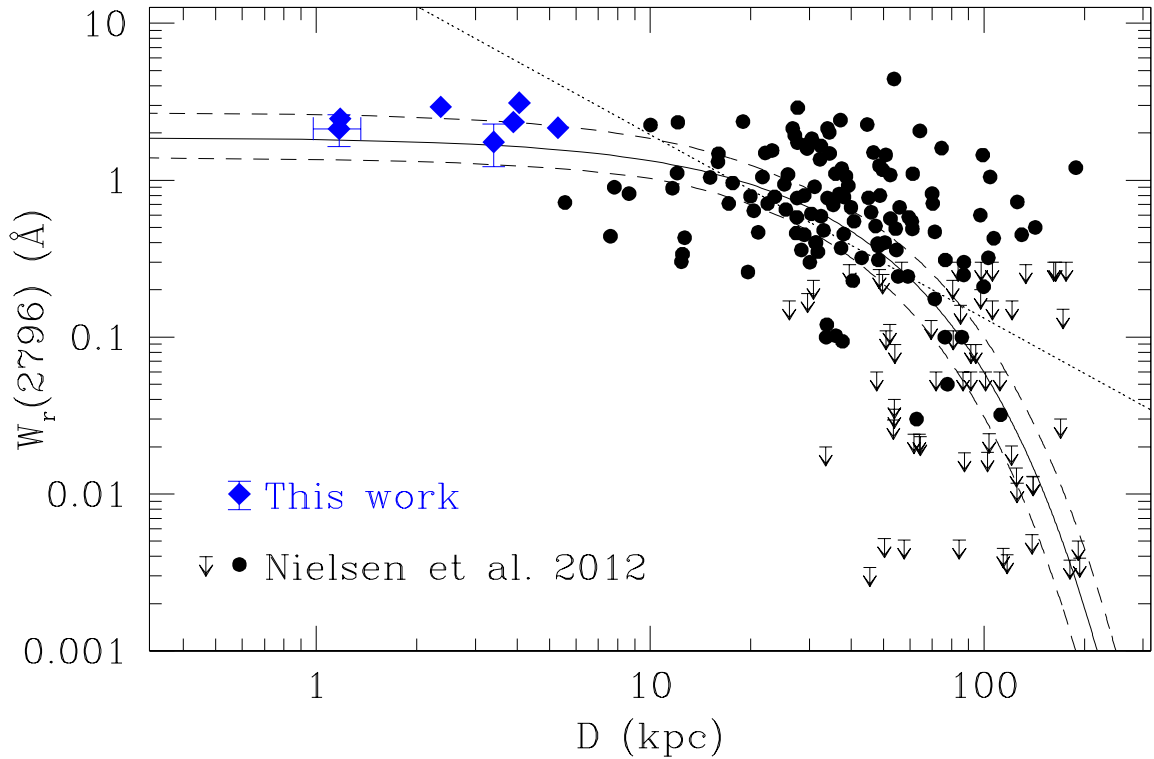}
\caption[angle=0]{The {\MgII} $\lambda$2796 rest-frame equivalent
  width, $W_r(2796)$, versus impact parameter, $D$. Galaxies with
  detected {\MgII} absorption are presented as filled point-types
  while those with upper limits are indicated with downward
  arrows. The circles and arrows represent the {\magiicat} sample
  comprising 182 galaxies and the solid line is a maximum likelihood
  log-linear fit where log$[W_r(2796)]=(- 0.015\pm0.002)\times
  D+(0.27\pm0.11)$ and the dashed curves provide 1~$\sigma$
  uncertainties \citep{nielsen12,nielsen13}. The dotted line is the
  power-law fit derived by \citet{chen10a}. Our new objects have
  $D<6$~kpc and are indicated with blue diamonds. 
}
\label{fig:main}
\end{center}
\end{figure*}

\section{Results \& Discussion}\label{sec:results}

In Table~\ref{tab:LRIS} we present seven new $z\sim0.1$ {\MgII}
absorption systems that have measured rest-frame equivalent widths in
the range of $1.75\leq W_r(2796)\leq 3.11$~{\AA}. The absorbers are
detected within $1.17\leq D\leq 5.30$~kpc of their host galaxies and
have relative velocity offsets of $10\leq\Delta v\leq 530$~{\kms}.  A
detailed kinematics study will be presented in an upcoming paper.  The
host galaxies have a $\left<L_R\right>=0.5L_{\star}$ and a
$\left<{\mbox{SFR}}\right>=2$~M$_{\odot}$yr$^{-1}$ \citep{york12}.
These systems represent the lowest impact parameters probing {\MgII}
absorption around known host galaxies. At these projected separations
we are likely probing {\HI} column densities of log[N(\HI)]$\sim
19-21$ that are typically detected around local star-forming galaxies
and extend out to 10--15~kpc \citep[see][]{sancisi08}.

In Figure~\ref{fig:main} we show the current state of the
$W_r(2796)-D$ relation for known {\MgII} absorption systems that are
associated with spectroscopically identified host galaxies. The
circles and limits represent a body of work from the literature
consisting of 182 absorbers and non-absorbers, respectively, recently
compiled in the
{\magiicat}\footnote{http://astronomy.nmsu.edu/cwc/Group/magiicat/}
\citep{nielsen12,nielsen13}. For our analysis, we have added an
additional {\magiicat} galaxy to our $D$<6~kpc sample that is
associated with a $W_r(2796)=0.72$~{\AA} absorber at $D=5.4$~kpc
\citep{steidel93,gb97}.  \citet{nielsen12} found a 7.9~$\sigma$
anti-correlation between $D$ and $W_r(2796)$ that is best fit by
log$[W_r(2796)]=(−-0.015\pm0.002)\times D+(0.27\pm0.11)$. The dotted
line is a power-law fit by \citet{chen10a} to a subset of 71 systems.

Our seven new absorbers are presented as diamonds (blue) in
Figure~\ref{fig:main}. We find that the \citet{chen10a} extrapolated
fit over-estimates the expected equivalent widths at low impact
parameters.  However, we find that the eight $D<6$~kpc absorbers
follow the {\magiicat} fit, even around $\sim$1~kpc. This is
interesting since {\MgII} absorption at higher impact parameters is
expected to trace a range of kinematically different, infalling,
outflowing, metal-poor, metal-rich gas structures, while gas within a
few kiloparsecs should be associated with an extended, infalling, and
co-rotating extra-planar disk \citep[e.g.,][]{fraternali02,heald07a}
combined with possible wind signatures
\citep[e.g.,][]{martin12,rubin13}.  Therefore it is interesting that,
to first order, the equivalent widths, which measure both gas
kinematics and quantity, exhibit a smooth continuity with the
log-linear fit based upon extended halo gas.  This may reflect recent
results that {\MgII} gaseous halos are self-similar when normalized by
the galaxy halo mass, indicating that regardless of the kinematic and
gas origins, the halo mass is the dominant driving factor in
determining their properties \citep{churchill13a,churchill13b}. The
smooth $W_r(2796)-D$ relation does not necessarily imply that the gas
distribution within an individual galaxy is smooth, but rather that
gas patchiness and kinematic/column density profiles conspire to yield
a monotonically decrease gas profile for a heterogeneous population of
galaxies with random orientations.

The halo gas covering fraction is found to be unity within 10~kpc,
which is consistent with the unity {\HI} emission covering fraction
for local star-forming galaxies out to 10~kpc
\citep[see][]{sancisi08}. The eight absorbers within 6~kpc have a mean
$\left<W_r(2796)\right>=2.2$~{\AA}; using the \citet{menard09}
N(\HI)--$W_r(2796)$ relation this translates to log[N(\HI)]$\sim20.1$,
the typical column density of DLAs/sub-DLAs at $D\sim7$~kpc of galaxy
disks \citep{zwaan05b}. In addition, the covering fraction remains
unity out to 25~kpc. Our eight galaxies are expected to contain a
significant {\HI} gas component since they were selected by emission
lines produced by ongoing star formation.

Given that the $W_r(2796)$ distribution is roughly constant within
10~kpc, with unity covering fraction, we can assume that the column
density profile is roughly constant towards the galaxy center with
$\left<\hbox{log[N(\HI)]}\right>\sim20.1$ (for
$\left<W_r(2796)\right>=2.2$~{\AA}). The product of the column density
and a gas cross-section of radius 10~kpc yields an estimated total
{\HI} mass traced by {\MgII} of $M_{\hbox{\tiny \HI}}=1.3\times
10^8$~M$_{\odot}$.  The eight galaxies in our analysis have a mean
luminosity of $\left<L_R\right>=0.5L_{\star}$ and are expected to have
a typical total {\HI} gas mass of $M_{\hbox{\tiny \HI}}\sim5\times
10^9-2\times10^{10}$~M$_{\odot}$ \citep{huang12}. Our derived {\HI}
mass is consistent with the expected halo/extra-planar {\HI} gas mass,
which are roughly 1--30\% of the total {\HI} gas mass
\citep{sancisi08}.  However, since the quasar lines-of-sight pass very
near to and through the galaxy disk, it is expected that the computed
{\HI} mass traced by {\MgII} absorption is equivalent to the total
expected {\HI} mass. This {\HI} mass inconsistency could be caused by:
(1) the N(\HI)--$W_r(2796)$ relation under-predicting the amount of
{\HI} at high $W_r(2796)$, (2) the {\MgII} absorption tracing only
halo/extra-planar material outside the higher column density disk, or
(3) the $W_r(2796)-D$ relation having a sharp turn-up in equivalent
width at $D$ smaller than we have probed with this work in order to
account for the remaining missing gas mass.

\citet{zwaan05b} reproduced the incidence rate of high redshift DLAs
assuming that DLAs arise from disk/ISM gas of local galaxies and found
that the median $D$ giving rise to DLAs was 7.3~kpc; we apply this
$D$-cut which is equivalent to a $D<6$~kpc cut for our study (see
Figure~\ref{fig:main}). To examine if the equivalent widths found for
the $D<6$~kpc sight-lines are representative of a combination of halo
gas and the interstellar medium (ISM), we use the best studied case --
the Milky Way (MW).  The most comprehensive absorption-line analysis
of the MW was performed by the Hubble Space Telescope Quasar
Absorption Line Key Project \citep{bahcall93}. A total of 83 quasars
were observed with the Faint Object Spectrograph G190H and G270H
gratings identifying 85 Galactic {\MgII} systems: 71 {\MgII} systems
are identified as being free of IGM {\Lya} contamination from high
redshift sources \citep{savage00}. \citet{savage00} identified 71
systems as being ISM$+$HVC/halo gas (see their Table 7, columns
8$+$10) and 21 systems as HVC/halo gas (see their Table 7, column 10).
We do not attempt to differentiate between absorption produced by HVCs
or halo gas, so we adopt the notation of ``MW ISM$+$halo'' and ``MW
halo'' absorption systems. Although this is the best sample to compare
to, there is one caveat: the MW sight-lines pass through approximately
half of the disk/halo.  Accounting for the highly saturated nature of
the MW absorption-lines, and assuming the gas velocity dispersion is
symmetric about the disk-plane, we multiply the equivalent widths by a
factor of two to emulate a full line-of-sight through the MW.

\begin{figure}[t]
\includegraphics[angle=0,scale=0.43]{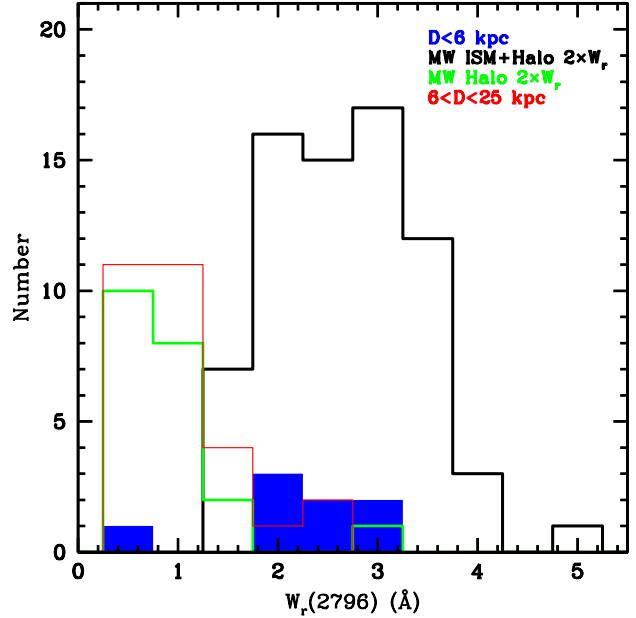}
\caption[angle=0]{The {\MgII} $\lambda$2796 rest-frame equivalent
  width, $W_r(2796)$, distributions are shown for the $D<6$~kpc (blue)
  and the $6<D<25$~kpc (red) quasar absorption-line systems. A
  comparison is made between intermediate redshift systems and those
  through the Milky Way (MW). A correction factor of two is applied
  since sight-lines are intercepted halfway though the disk/halo. The
  $W_r(2796)$ distributions of the the MW ISM+halo (black) and the
  halo (green) are shown. Note the $D<6$~kpc systems and those of the
  MW ISM+halo systems have similar distributions. The $6<D<25$~kpc
  systems have a similar distribution to the MW halo.  }
\label{fig:main2}
\end{figure}

In Figure~\ref{fig:main2}, we plot the equivalent width distributions
of the $D<6$~kpc sample and MW ISM+halo sample.  Note that the
$D<6$~kpc and the MW ISM+halo samples are similar with median values
of 2.25~{\AA} and 2.70~{\AA} and average values of 2.20~{\AA} and
2.70~{\AA}, respectively. A Kolmogorov-Smirnov (KS) test indicates
that the $D<6$~kpc and the MW ISM+halo $W_r(2796)$ distributions are
likely drawn from the same parent population with $P(KS)=0.576$
(0.80~$\sigma$). This implies that the $D<6$~kpc sample is likely
detecting the expected total ISM+halo of their host galaxies.  Given
that the MW ISM$+$halo $W_r(2796)$ distribution is consistent with the
$D<6$~kpc distribution, and that the ISM will unlikely produce higher
$W_r(2796)$ systems, we would not expect to observe a sharp upturn in
the $W_r(2796)-D$ relation as one approaches $D=0$~kpc. Although the
sight-lines through the MW intercept the disk at $D=8.5$~kpc, the
{\HI} radial column density profile is rather flat for $D<8.5$~kpc
\citep{kalberla09} and the $W_r(2796)$ distribution would likely be
similar if we were located elsewhere in the disk.

In Figure~\ref{fig:main2} we show the $6<D<25$~kpc equivalent width
distribution for intermediate redshift galaxies from {\magiicat}. Note
that compared to the $D<6$~kpc sight-lines, they appear quite
different with an average $W_r(2796)$ of 1.0~{\AA} compared to
2.2~{\AA} ($D<6$~kpc). A KS test shows that two populations differ at
the 3.1~$\sigma$ level ($P(KS)=0.002$). We further show the
$W_r(2796)$ distribution of MW halo systems in Figure~\ref{fig:main2},
which has an average $W_r(2796)$ of 0.86~{\AA}. The MW halo
distribution is similar to the $6<D<25$~kpc distribution with
$P(KS)=0.767$ (1.19~$\sigma$) and very different from the $D<6$~kpc
distribition at the 3.4~$\sigma$ level ($P(KS)=0.00072$). The similar
distribution between the MW halo and $6<D<25$~kpc systems suggest that
higher impact parameter quasar sight-lines systems likely probe halo
gas material at much lower {\HI} column densities.  Thus, small-$D$
{\MgII} absorption systems are indeed tracing the ISM or ISM$+$halo
gas from their host galaxies, relative to expectations from the MW.

Our initial {\HI} mass calculation must suffer from geometry and/or
filling factor assumptions and/or uncertainties of the
N(\HI)--$W_r(2796)$ relation \citep{menard09}. The N(\HI)--$W_r(2796)$
relation contains scatter over 4 orders of magnitude and DLA column
densities exist only for $W_r(2796)>$3~{\AA}, while DLAs are observed
in systems as low as $0.6$~{\AA} \citep{rao06}. Although this relation
provides an indirect means for computing the median N(\HI), it may
have limitations.

\section{Conclusion}\label{sec:conclusion}

We have obtained the first measurements of {\MgII} absorption
associated with spectroscopically confirmed star-forming galaxies at
projected distances of $D<6$~kpc. We find the following:

\begin{enumerate}

\item The well known anti-correlation between rest-frame {\MgII}
  equivalent width and impact parameter is maintained for systems with
  $0.5<D<6$~kpc and the gaseous halos have unity covering fraction
  within $D<25$~kpc. Furthermore, the $W_r(2796)$ relation converges
  to $\sim2$~{\AA} at $D=0$~kpc.

\item When compared to the Milky Way, we find that our $D<6$~kpc
  sample is consistent with the $W_r(2796)$ distributions of the MW
  ISM+halo; the majority of the $W_r(2796)$ is produced by the
  ISM. This indicates that our new systems are likely probing the
  ISM/disk and halo gas of their host galaxies.

\item The comparison between the MW and small-$D$ absorption-line
  systems indicates that the halo gas profile likely decreases
  monotonically from the galaxy center out to $\sim150$~kpc, with the
  covering fraction decreasing as a function of $D$. This implies no
  rapid increase in $W_r(2796)$ at $D$ approaches 0~kpc. Our data show
  there is a smooth log-linear continuity in the gas content from the
  outer halo all the way to the halo/extra-planar/ISM interface of
  galaxies and is not in agreement with the extrapolations of
  power-law fit of \citep{chen10a}.

\item When compared to the MW, quasar sight-lines with $6<D<25$~kpc
  have a $W_r(2796)$ distribution similar to that of MW halos,
  implying that beyond $D\sim6$~kpc, quasar sight-lines likely are
  tracing halo gas only.

\end{enumerate}

It is remarkable that gas profiles of galaxies can be fit by a single
log-linear $W_r(2796)-D$ relation over such large scales and range of
gas-phase conditions. Consistent with the MW, this relation is
maintained through the halo/extra-planar/ISM interfaces regardless of
their complex kinematic interplay. This further implies a smooth {\HI}
column density profile surrounding galaxies below current
observational limits.

These low redshift, small-$D$ absorption systems will be important in
bridging the gap between absorption-line studies and future {\HI}
observations/surveys, which could play a key role in understanding the
CGM.


ERW acknowledges Australian Research Council grant DP 1095600. CWC was
partially supported through the Research Enhancement Program from
NASA's New Mexico Space Grant Consortium.  NMN was partially support
through an NSF EAPSI grant and a GREG at NMSU.  Data was obtained at
the W.M. Keck Observatory, which is operated as a scientific
partnership among the California Institute of Technology, the
University of California and the National Aeronautics and Space
Administration. The Observatory was made possible by the generous
financial support of the W.M. Keck Foundation.  Data was also obtained
from The Sloan Digital Sky Survey (SDSS/SDSS-II), which is funded by
the Alfred P. Sloan Foundation, Participating Institutions, NSF,
U.S. Department of Energy, NASA, Japanese Monbukagakusho, Max Planck
Society, and the Higher Education Funding Council for England.





{\it Facilities:} \facility{Keck I (LRIS)}, \facility{Sloan (SDSS)}.

\end{document}